# Rapid accumulation of colloidal microspheres flowing over microfabricated barriers

P. Prakash[1,a)], A. Z. Abdulla[2,b)], M. Varma[1,3,*]

[1]Centre for Nanoscience and Engineering, Indian Institute of Science, Bangalore, 560012, India
[2]Department of Physics, Indian Institute of Science, Bangalore, 560012, India
[3]Robert Bosch Centre for Cyber Physical Systems, Indian Institute of Science, Bangalore, 560012, India
Author to whom correspondence should be addressed: [*]**mvarma@iisc.ac.in**

Accumulation of particles while flowing past constrictions is a ubiquitous phenomenon observed in diverse systems. Some of the common examples are jamming of salt crystals near the orifice of salt shakers, clogging of filter systems, gridlock in traffics etc. For controlled studies, accumulation events are often examined as clogging process in microfluidic channels. Experimental studies thus far have provided with various physical insights, however, they fail to address commonly encountered accumulation events relevant to human health such as dental and arterial plaques. We simulate arterial plaque like accumulation events by flowing colloidal microspheres over micro-structured barriers in microfluidic environment. Our experiments reveal the role of electrostatic, contact and hydrodynamic forces in facilitating plaque-like build up events. A decrease in Debye length (electrostatic repulsion) between interacting surface by two orders leads to only a minor increase in accumulation. In contrast, an increase in the roughness by 3 times results in dramatic rise of accumulation.

In contrast to the clogging events [1], the accumulation processes in arteries are gradual forming heaps of cellular milieu over time [2]. Such accumulation process usually entails simultaneous action of DLVO, contact and hydrodynamic forces [3]. The dominant physical force promoting accumulation usually is system dependent. For example, in control settings, aggregation in colloid is driven by DLVO forces [4], jamming events are supported by contact forces [5], whereas formation of active crystals is entirely due to hydrodynamic interactions [6,7]. The real accumulation processes however, involves multiple physical phenomena making it exceptionally challenging to decipher dominant mechanism.

---

[a)] Present address: Warwick Integrative Synthetic Biology Centre, University of Warwick, Coventry, CV4 7AL, United Kingdom.
[b)] Present address: Laboratoire de Biologie et Modelisation de la Cellule, ENS de Lyon, 69364 Lyon Cedex 07, France.

In addition, there can be often ignored phenomena such as non-laminar flow, rheology of colloidal suspension, tribological effects etc. which can further increase the complexity [8–10]. Colloidal systems have long been used to decipher several aspects of the accumulation process. However, to meaningfully understand process of arterial blockage, we need to mimic certain aspect of arterial geometry which facilitates the gradual build-up of plaque. The arterial blockage is initiated by necrotic core which essentially is heap of the dead cells. We imitate the necrotic core by microfabricating glass substrate in triangular shape similar to the necrotic core [11] whereas in place of cells we use polystyrene microspheres. By using the microfabricated substrate in controlled environment we study the role of electrostatic repulsion and contact forces in facilitating plaque-like build up events. In this manuscript, we have discussed various aspects of the accumulation process and its modelling.

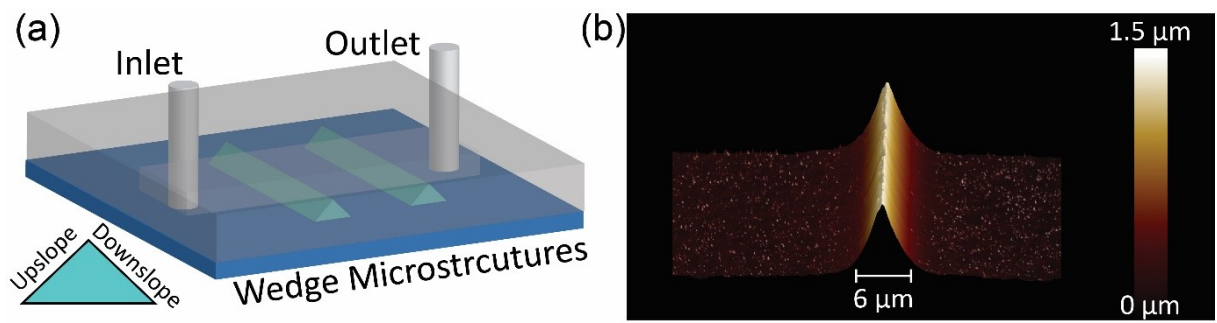

**Fig. 1** (a) Schematic of PDMS microfluidic device (h = 100 μm, w = 500 μm) affixed over wedge microstructures. (b) AFM of microstructures (h = 1.5 μm, w = 6 μm).

To understand the process of plaque-like build up, we studied accumulation of microspheres flowing over wedge-like fabricated structures in microfluidic environment (Fig. 1(a)). The wedge microstructures are fabricated by patterning 2.2 μm thick and 5 μm wide photoresist (S1813) in an array with a gap of 50 μm on glass substrates. Subsequently, resist patterned substrate is isotopically etched for 15 sec in HF etchant (10% HF + 7% HCl) which yields wedge shaped symmetric microstructures of height ~1 μm and width ~6 μm (Fig. 1(b)) [12]. A PDMS (Polydimethylsiloxane) microfluidic device of height 100 μm and width 500 μm is clamped on the microstructured substrate to perform experiments. The device is imaged by a custom-built inverted CMOS camera assembly which is illuminated using ring light. The colloidal microspheres of size 10 μm (0.5% w/v) in deionized ultra-filtered water (DI water) selectively pin on the downslope side of microstructures when passed at a flow rate of 10 μl/min (Fig. 2(a), SI Video 1). The pinning of microspheres in downslope region is observed for flow rates as low as 0.1 μl/min. For even lower flow rates, microspheres either do not pin or tend to pin in the upslope region due to their inability of crossing upslope gravitational potential barrier. No pinning is observed at higher flow rates beyond 50 μl/min. The horizontal drag force on microspheres at a flow rate of 10 μl/min is 0.12 nN regardless of pinning in upslope or downslope region. However, the microspheres experience a lift force of 11.7 pN when pinned on upslope region as opposed to a downward thrust of −13.5 pN when pinned on downslope region which enhances the contact force (See SI Sec. 1). The same 10 μm microspheres when flown in 0.01 M phosphate buffer saline (PBS), forms a linear chain of colloidal microspheres as shown in Fig. 2(b). The chain formation proceeds by pinning of single microspheres in the downslope region which then act as seed for successive microspheres to deposit one by one (See SI Video 2). As seen in the video, close contact between incoming and already pinned microspheres is crucial for the microspheres to glide their way and form linear chains.

The interplay among hydrodynamics, DLVO ($F_{DLVO}$) and contact forces are responsible for this fascinating rearrangement led linear alignment of microspheres. Noticeably, such linear arrangements are possible only in the presence of 0.01 M PBS which indicates the role of salt solution in suppressing electrostatic repulsion. The DLVO force comprises of electrostatic double layer repulsion ($F_{EDL}$) and attractive Van der Waals force ($F_{VDW}$). The zeta potential of microspheres drops from -31.5 mv in DI water to -4.33 mv in 0.01 M PBS and the corresponding Debye length ($\sim 0.3/\sqrt{[C]}$ nm, C in M) drops from 948.7 nm in DI water to 3 nm respectively [13]. The decrease in electrostatic potential and Debye length in PBS medium leads to lower repulsion ($F_{EDL}$) and higher chances of contact allowing microspheres to graze closely resulting in serial accumulation. Further, the attractive Van der Waals component of DLVO force relies on the gap between microspheres and substrate which is set by the roughness of interacting surfaces [14]. The roughness of glass substrate is $R_a \sim 8.0$ nm (SI Fig. 2(a)), setting a lower limit on the minimum distance of interaction with microspheres. The DLVO interaction force between microsphere and glass substrate is attractive only at sub-nm length scales as shown in Fig. 2(c) (See SI Sec. 2). Therefore, we conclude that the microspheres are predominantly held by frictional contact force in the presence of flow.

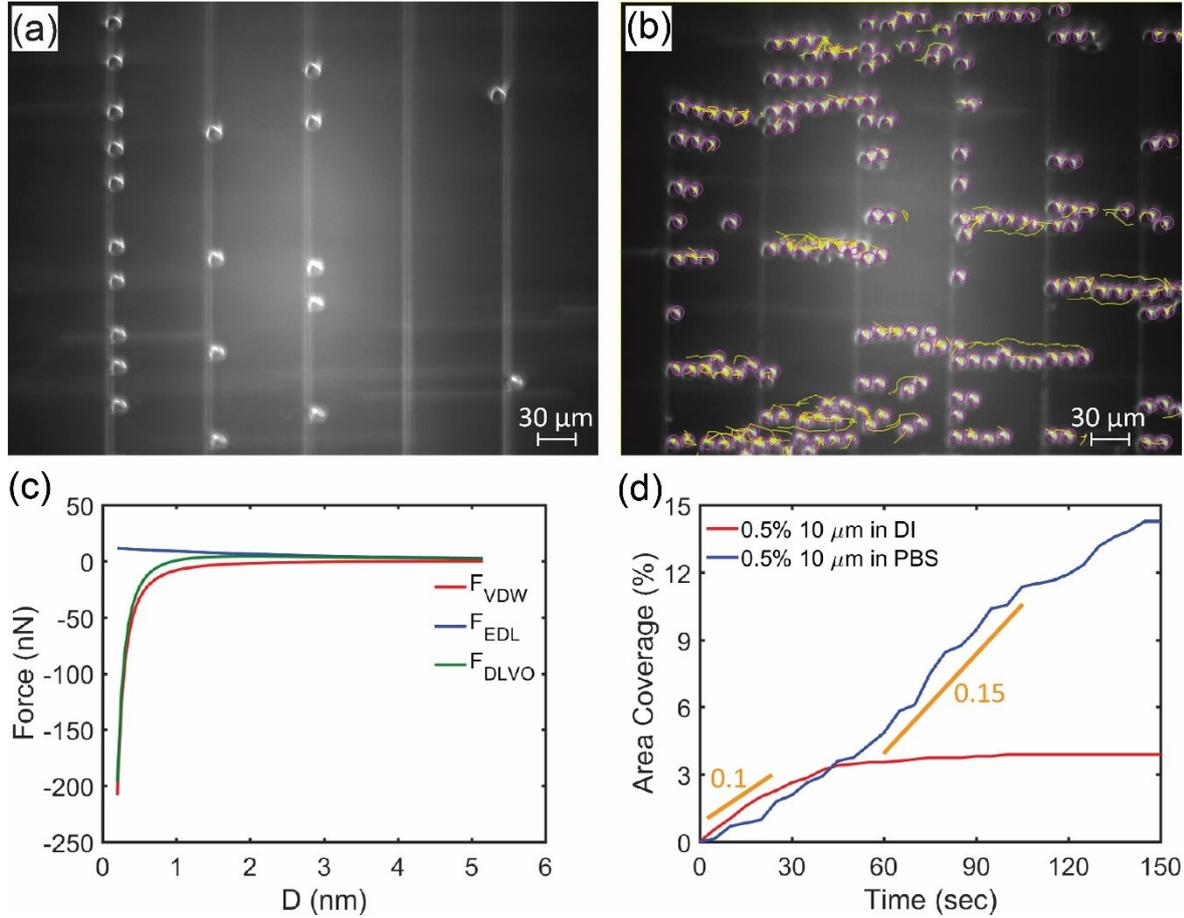

**Fig. 2** Accumulation of 10 μm microspheres at a flow rate of 10 μl min$^{-1}$. (a) Microspheres in the presence of DI water pin on downslope region of wedge microstructures. (b) Microspheres in the presence of 0.01 M PBS accumulate in the form of linear chains. (C) DLVO force between microsphere and plain glass substrate is attractive for sub-nm length scale. (d) Area of coverage in the presence of DI water saturates as the downslope sites start to fill over time whereas continuously increase due to formation of linear chains in PBS medium.

The stick and slip motion exhibited by microspheres as seen in SI Video 2 is a characteristic feature of frictional contact force ($F_{CON}$) and we estimate it as follows:

$$F_{CON} = F_{ST} - F_V$$

Where, $F_{ST}$ is the drag force on a static microsphere in the presence of flow and $F_V$ is the drag force when microspheres are slipping against the pinned microspheres at a speed 'V'. The microspheres are flown in the device at a flow rate of 10 μl/min which yields $F_{ST} = 0.12$ nN as estimated from simulations. Drag force near a surface can be analytically estimated as $1.7 \times F_{Stokes}$, where $F_{Stokes} = 6\pi\eta rV$ is the stokes drag on a sphere slipping with speed 'V', where $\eta = 8.9 \times 10^{-4}$ Pa.s is the viscosity of water and r = 5 μm is the radius of microspheres [15]. The average slip velocity 'V = 68.8 μm/s' is estimated from the video analysis of yellow tracks in Fig. 2(b) which gives $F_V = 9.8$ pN. Understandably, the contact force $F_{CON} \cong 0.12 \text{ nN} - 9.8 \text{ pN}$ is marginally lower than the hydrodynamic drag force on static microspheres which is essential for such rearrangements to occur. The percentage area coverage is much higher in PNS medium due to linear chain formation (Fig. 2(d)). Over time the area coverage saturates in DI water which is due to filling of pinning sites whereas in the presence of PBS, the already pinned microspheres act as a seed for the formation of linear chains eventually covering more than 80 % of the total area (Fig. SI 2(b)). The slope of area coverage curve as shown in Fig. 2(d) is expected to be linear as it depends only on the number of microspheres flowing per unit time which is determined by the flow rate of solution.

The increased accumulation due to formation of linear chains is possible due to use of PBS which screens the electrostatic repulsion thereby increasing the chance of contact between microspheres. Alternatively, even in the presence of DI water the probability of contact can be increased by increasing the flow rate which inevitably also increases the drag force but can be compensated by enhancing roughness. We sourced 4 µm microspheres of roughness 1.5 nm (Fig. 3(a)) and 4.7 nm (Fig. 3(b)) respectively (See SI Sec. 3). The pinning of 10 µm microspheres similar to Fig. 2(a) is observed when a mixture of 0.25% w/v 10 µm and 0.1% w/v 4 µm microspheres of lower roughness ($R_a$ = 1.5 nm) is flown (10 − 70 µl/min) with DI water. In contrast, a rapid accumulation of microspheres is observed (SI Video 3) when 0.25% w/v 10 µm microspheres are flown (10 − 70 µl/min) with 0.1% w/v 4 µm rougher microspheres ($R_a$ = 4.7 nm, Fig. 3(b)) in DI water. Initially, microspheres are flown at the rate of 10 µl/min which pins microsphere of size 10 µm in the downslope region. The pinned microspheres then act as seed for rapid accumulation at higher flow rate of 70 µl/min. The rapid accumulation proceeds by pinning of 4 µm rougher microspheres on already pinned 10 µm microspheres where, incoming 10 µm microspheres latch onto 4 µm microspheres and the process continues. Interestingly, the Zeta potential -32.38 mv of 4 µm rougher microspheres ($R_a$ = 4.7 nm) is higher than that of smoother microspheres ($R_a$ = 1.5 nm) -26.39 mv. So, the build-up of rougher microsphere facing higher electrostatic repulsion indicates the dominant role of contact forces in accumulation. The accumulation in multi microsphere (10 + 4 µm in DI water) system grows in 3-dimension (Fig. 3(c)) in contrast to 1-dimensional (10 µm in PBS) linear chains previously discussed (Fig. 2(b)). This results in a dramatic increase of total area coverage to 30 % (Fig. 3(d)) in just one minute as compared to only 5 % in linear chain forming events (Fig.2(d)).

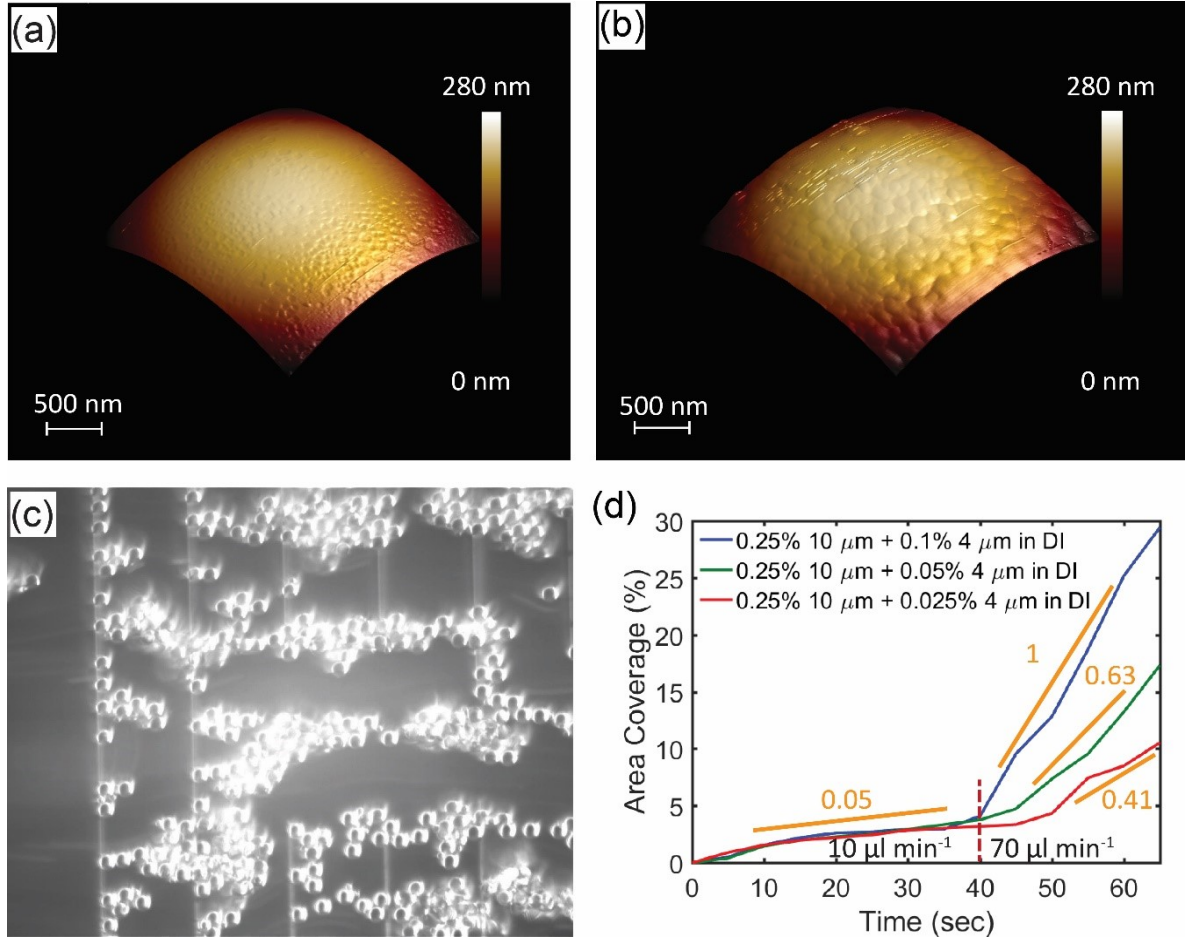

**Fig. 3** Accumulation of mixed microspheres (4 μm + 10 μm) over wedge microstructures. (a) AFM of 4 μm smoother microspheres of roughness $R_a$ = 1.5 nm. (b) AFM of 4 μm rough microspheres ($R_a$ = 4.7 nm). (c) Rapid accumulation of microspheres (4 μm + 10 μm) at a high flow rate of 70 μl min$^{-1}$. (d) The rate of AOC jumps 20-fold as the flow rate is switched from 10 μl/min to 70 μl/min.

The area of coverage (AOC) in accumulation process is linear with time, indicating accumulation to be a single particle depositing one by one as they arrive. Understandably, the rate of single pinning events of 10 μm microspheres decreases from 0.1 to 0.05 as the concentration of microspheres is reduced from 0.5 % in Fig. 2(d) to 0.25 % in Fig. 3(d). Hence, we empirically define the area of coverage (AOC) in time 't' as:

$$AOC = SP \times [C] \times t$$

Where, 'SP' is the sticking probability and '[C]' is the concentration of 10 μm microspheres. The magnitude of sticking probability (SP) is expected to be a function of DLVO interaction, contact force and flow rate. If the flow rate is kept constant, then the sticking probability will essentially rely on material properties such as salt concentration and roughness. For instance, the slope of AOC vs time increases from 0.1 in DI water to 0.15 in PBS (Fig. 2(d)) where, the increase can be attributed to reduction in Debye length leading to an increased sticking probability. In the case of multi-sized microspheres (10 µm + 4 µm) in DI water, the accumulation rate is even higher where, as the flow rate is increased to 70 µl/min the rate increases 20 folds as shown in Fig. 3(d). Interestingly, drop in the rate of AOC isn't linear with the reduction in concentration of 4 µm microspheres and instead reduces by ~60% with 50% reduction in the concentration.

To understand the gradual, build-up process we closely examined the individual accumulation events (SI Video 4). The rapid build-up of microspheres at flow rate of 70 µl/min is initiated by pining of 4 µm microspheres over 10 µm microspheres. The vectorial flow over pinned 10 µm microsphere is shown in Fig. 4(a) where, we see the bulk of flow is directed at position 3-4 (Fig. 4(b)). This suggests the preferred initial pinning of 4 µm microspheres around position 3-4 which is also seen in the experimental video. The drag force is maximum around position 3-4 which may seem undesirable for pinning, however, high drag force also increases the probability of contact suggesting the pinning position of smaller microspheres is dominated by chances of initial contact. This is in contrast with the pinning of successive 10 µm microspheres which undergoes stick and slip motion before settling on the most stable position. It can be seen during many instances that the 10 µm microsphere glides its way towards the posterior region after initial contact with 4 µm microsphere. The vectorial flow (Fig. 4(c)) guides the microsphere after initial contact with 4 µm microsphere on the top where drag force is 1.25 nN to the posterior region where it reduces to ~0.49 nN (Fig. 4(d)).

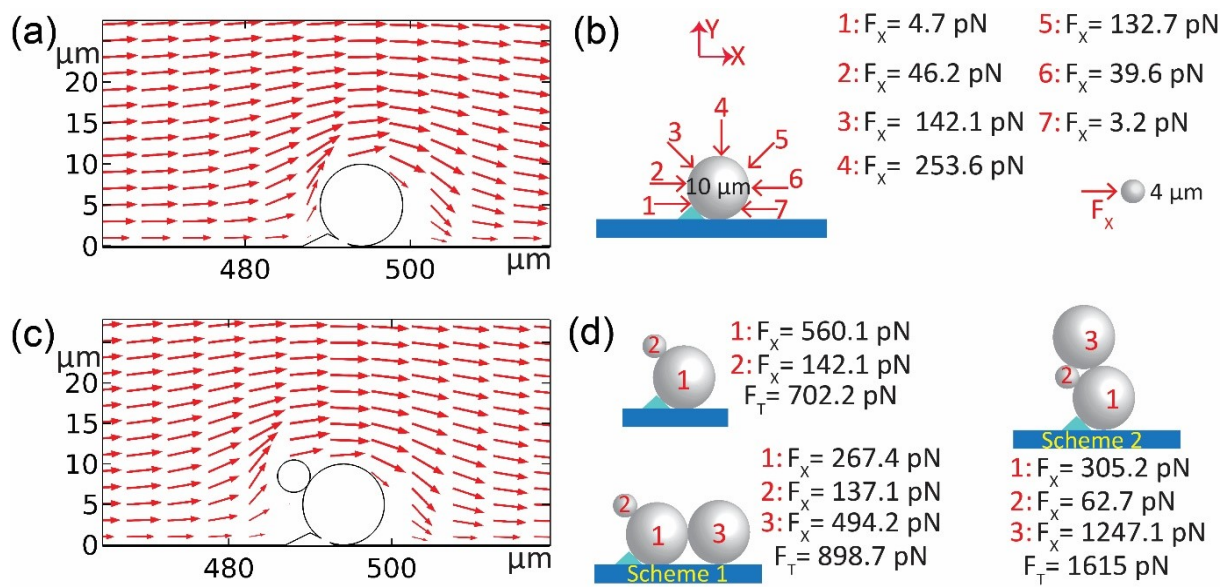

**Fig. 4** Hydrodynamic drag on microspheres at different positions. (a) Vectorial direction of flow over 10 µm microspheres. (b) The hydrodynamic drag force on 4 µm microsphere at different positions.

Surprisingly, positions around 3 facing maximum hydrodynamic drag is the preferred position of pinning as seen from the videos. (c) Vectorial direction of flow over 10 μm and 4 μm microsphere system. (d) Increase in drag force due to sticking of incoming 10 μm microsphere on 10 μm and 4 μm microsphere system. Microspheres face higher drag in Scheme 2 and usually glide to form Scheme 1 configuration which faces much lower drag.

To summarize, the accumulation process is initiated by pinning of 10 μm microspheres on downslope region when flown (10 μl/min) over wedge shaped symmetric microstructures of height 1.5 μm and width 6 μm. The incoming microspheres interact with the pinned microspheres through DLVO and contact forces. The electrostatic component of DLVO force can be decreased by introducing salt solution such as PBS buffer (0.01 M) which leads to the formation of linear chains in the presence of 10 μm microspheres (Fig. 5(a)). Alternatively, at higher flow rates (70 μl/min) 4 μm microspheres of roughness $R_a = 4.7$ nm is introduced which act as a bridge between incoming 10 μm microspheres and already pinned 10 μm microspheres (Fig. 5(b)) resulting in rapid accumulation.

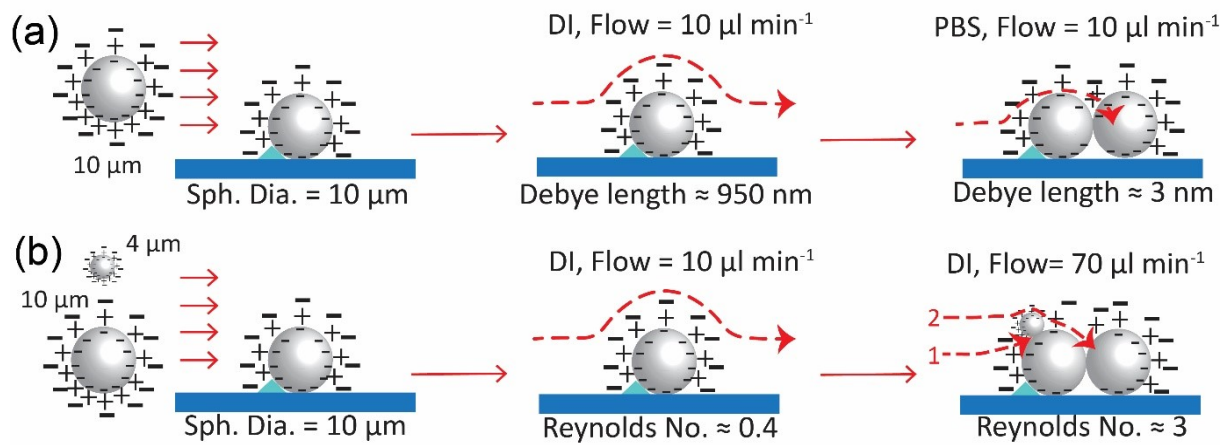

**Fig. 5** Schematic showing possible route towards accumulation process. (a) Screening of Electric double layer force by use of PBS enhances the probability of contact between microspheres thereby assisting in formation of linear chains. (b) Rougher 4 μm microspheres at high flow rate pin onto 10 μm microspheres starting off rapid accumulation of incoming microsphere.

The accumulation events of colloidal microspheres are reminiscent of Myocardial Infarction commonly known as heart attack occurs due to blockage of arteries which is caused by gradual build-up of plaque [16]. Further, the experimental condition of our system is similar to that in blood vessels. For instance, the size of microspheres (10 μm + 4 μm) which we have used is very similar to the size of cellular milieu (Neutrophils: 12 − 14 μm, Platelets: 2 − 3 μm, R.B.C.: 6 − 8 μm) in arteries. Though, the Reynolds number of blood flow varies in arteries some portions indeed have a flow with Reynolds number around ~3. The accumulation of colloids can be initiated by microstructures as small as 5 μm in thickness which surprisingly lies within the lower limit of experimentally reported necrotic core thickness (4 μm) [17]. Importantly, a little increase in roughness from 1.7 nm to 4.7 nm leads to rapid accumulation of colloidal microspheres. Such results are very illuminating as they put bounds on the requirement of physical conditions such as roughness and size variation in cells that can lead to rapid plaque build up. Our work clearly lays out the chemical and physical condition required for rapid accumulation of colloidal microspheres giving useful insights into plaque-like accumulation phenomena.

**Reference:**

[1]  H. M. Wyss, D. L. Blair, J. F. Morris, H. A. Stone, and D. A. Weitz, Phys. Rev. E - Stat. Nonlinear, Soft Matter Phys. **74**, (2006).

[2]  P. Libby, P. M. Ridker, and G. K. Hansson, Nature **473**, 317 (2011).

[3]  E. Dressaire and A. Sauret, Soft Matter **13**, 37 (2017).

[4]  A. N. Radhakrishnan, M. P. C. Marques, M. J. Davies, B. O'Sullivan, D. G. Bracewell, and N. Szita, Lab Chip **18**, 585 (2018).

[5]  A. Donev, S. Torquato, F. H. Stillinger, and R. Connelly, J. Appl. Phys. **95**, 989 (2004).

[6]  J. Palacci, S. Sacanna, A. P. Steinberg, D. J. Pine, and P. M. Chaikin, Science (80-. ). **339**, 936 (2013).

[7]  R. Singh and R. Adhikari, Phys. Rev. Lett. **117**, (2016).

[8]  G. R. Wang, F. Yang, and W. Zhao, Lab Chip **14**, 1452 (2014).

[9]  M. Mohtaschemi, A. Puisto, X. Illa, and M. J. Alava, Soft Matter **10**, 2971 (2014).

[10] C. P. Hsu, S. N. Ramakrishna, M. Zanini, N. D. Spencer, and L. Isa, Proc. Natl. Acad. Sci. U. S. A. **115**, 5117 (2018).

[11] R. Hajhosseiny, T. S. Bahaei, C. Prieto, and R. M. Botnar, Arterioscler. Thromb. Vasc. Biol. **39**, 569 (2019).

[12] P. Prakash and M. Varma, Sci. Rep. **7**, 15754 (2017).

[13]　J. Israelachvili, *Intermolecular and Surface Forces* (Elsevier Inc., 2011).

[14]　S. Bhattacharjee, C.-H. Ko, and M. Elimelech, *DLVO Interaction between Rough Surfaces* (1998).

[15]　S. P. Mulvaney, C. L. Cole, M. D. Kniller, M. Malito, C. R. Tamanaha, J. C. Rife, M. W. Stanton, and L. J. Whitman, Biosens. Bioelectron. **23**, 191 (2007).

[16]　Y. Kojima, J. P. Volkmer, K. McKenna, M. Civelek, A. J. Lusis, C. L. Miller, D. Direnzo, V. Nanda, J. Ye, A. J. Connolly, E. E. Schadt, T. Quertermous, P. Betancur, L. Maegdefessel, L. P. Matic, U. Hedin, I. L. Weissman, and N. J. Leeper, Nature **536**, 86 (2016).

[17]　A. P. Burke, A. Farb, G. T. Malcom, Y. Liang, J. Smialek, and R. Virmani, N. Engl. J. Med. **336**, 1276 (1997).

## 1. Hydrodynamic equations for COMSOL simulation:

The force on microspheres is calculated by setting up hydrodynamics equations in accordance to channel geometry and size. The Navier-Stokes equation (Eq. 1) together with continuity condition ($\nabla \cdot u = 0$) for an incompressible fluid ($d\rho/dt = 0$) determines the flow.

$$\rho \frac{\partial u(t)}{\partial t} + \rho(u \cdot \nabla u) = -\nabla p + \mu \nabla^2 u \qquad (1)$$

Where, u is the velocity and $\rho, \mu$ are density and viscosity of water respectively. No-slip boundary condition ($u = 0$) is applied on all surfaces (Fig. 1(a)).

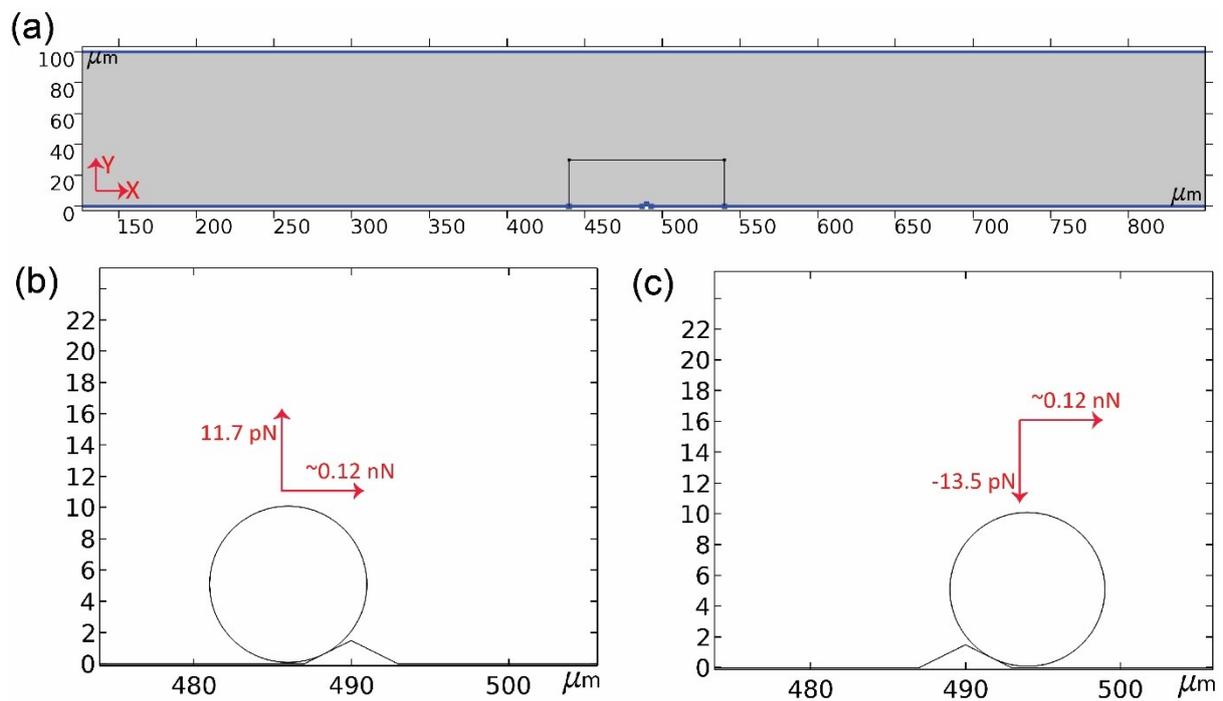

SI Fig. 1 FEM simulation geometry. (a) Microfluidic channel having a height of 100 μm with wedge microstructures in the centre. (b) Forces on 10 μm microsphere pinned in the upslope region. (c) Forces on 10 μm microsphere pinned in the downslope region.

## 2. DLVO force calculation

The electrostatic double layer repulsion ($F_{EDL}$) and attractive Van der Waals force ($F_{VDW}$) are together known as DLVO force ($F_{DLVO} = F_{EDL} + F_{VDW}$). The Van der Waals force between a spherical particle and flat surface is defined as [1]:

$$F_{VDW} = -AR/6D^2 \qquad (2)$$

Where, $A \approx 1 \times 10^{-20}$ J is the Hamaker constant, R = 5 µm is the radius of microsphere and D is the distance between particle and flat surface. The electrostatic double layer force between a spherical particle and flat surface is [1]:

$$F_{EDL} = KRZe^{-KD} \qquad (2)$$

Where, $K^{-1} = 0.3/\sqrt{[C]}$ nm is the Debye length for a monovalent electrolyte (major constituent of PBS buffer is NaCl which is monovalent), R = 5 µm is the radius of microsphere and D is the distance between particle and flat surface. $Z = 9.22 \times 10^{-11} \tanh^2(\psi_o/103)$ Jm$^{-1}$ is the interaction constant and $\psi_o \approx -30$ mv is surface potential [2].

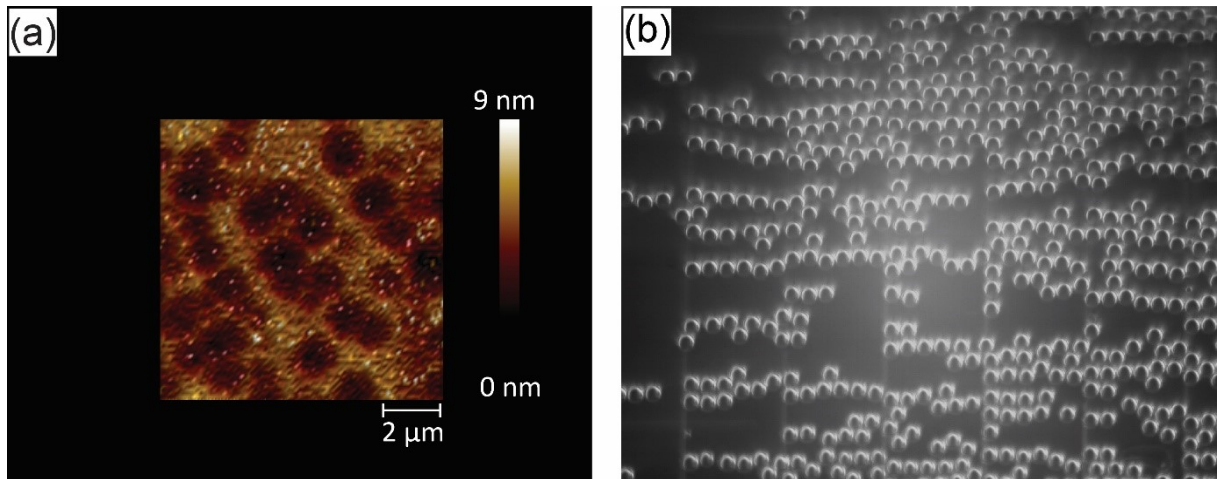

**SI Fig. 2** (a) AFM of glass substrate reveals its roughness as ~8.0 nm. (b) Linear chains of 10 µm microspheres eventually (~10 min) covering 80 % of the total area.

## 3. Materials and methods

Photoresist – S1813, Camera – Thorlabs, DCC1545M; Light source - EO, 63-306; Microspheres of size 10 µm – Sigma, 72986-5ML-F; Smooth microspheres of size 4 µm – Sigma, 81494-5ML-F; Rough microspheres of size 4 µm – Spherotech, PP-40-10.

**References:**

[1] J. Israelachvili, *Intermolecular and Surface Forces* (Elsevier Inc., 2011).

[2] D. S. Wright, B. S. Flavel, and J. S. Quinton, in *Proc. 2006 Int. Conf. Nanosci. Nanotechnology, ICONN* (2006), pp. 634–636.